# CLIMATE CHANGE VIA CO$_2$ DRAWDOWN FROM ASTROPHYSICALLY INITIATED ATMOSPHERIC IONIZATION?

Adrian L. Melott[a], Brian C. Thomas[b], and Brian D. Fields

**Abstract:** Motivated by the occurrence of a moderately nearby supernova near the beginning of the Pleistocene, possibly as part of a long-term series beginning in the Miocene, we investigate whether nitrate rainout resulting from the atmospheric ionization of enhanced cosmic ray flux could have, through its fertilizer effect, initiated carbon dioxide drawdown. Such a drawdown could possibly reduce the greenhouse effect and induce the climate change that led to the Pleistocene glaciations. We estimate that the nitrogen flux enhancement onto the surface from an event at 50 pc would be of order 10%, probably too small for dramatic changes. We estimate deposition of iron (another potential fertilizer) and find it is also too small to be significant. There are also competing effects of opposite sign, including muon irradiation and reduction in photosynthetic yield caused by UV increase from stratospheric ozone layer depletion, leading to an ambiguous result. However, if the atmospheric ionization induces a large increase in the frequency of lightning, as argued elsewhere, the amount of nitrate synthesis should be much larger, dominate over the other effects, and induce the climate change. More work needs to be done to clarify effects on lightning frequency.

Key Words: supernovae, glaciation, climate change

a Department of Physics and Astronomy, University of Kansas, Lawrence, KS 66045 USA. Email: melott@ku.edu

b Department of Physics and Astronomy, Washburn University, Topeka, Kansas 66621 USA. Email: brian.thomas@washburn.edu

c Department of Astronomy and Department of Physics, University of Illinois, Urbana, IL 61801 USA. Email: bdfields@illinois.edu

# 1. Introduction

There are various effects from relatively nearby supernovae, beyond the traditional emphasis on ozone depletion due to the ionization of the Earth's atmosphere. These include muon irradiation on the ground, the fertilizer effect of nitrates generated in the atmosphere, the fertilizer effect of iron deposited from the supernova, and indirect second-order effect of nitrate deposition from lightning, which is itself enhanced by atmospheric ionization. We have found that many of them are too small to be important, and others in fact competitive in intensity to ozone depletion.

In this paper we explore the idea that deposition of nitrate on the biosphere from the effects of ionizing radiation from a moderately nearby (50 pc) supernova might provide enough fertilizer effect to cause a drawdown of $CO_2$ and thereby cool the climate, possibly setting off the glaciations typical of the Pleistocene (as well as possibly at the end of the Ordovician). This is interesting because the Plio-Pleistocene transition to an icehouse climate is close to the time indicated by cosmogenic nuclide deposits for one or more nearby supernovae. The production of $NO_2$ results from the ionization and dissociation of molecular nitrogen, the resulting species reacting with oxygen in the atmosphere. The $NO_2$ may react with water or hydroxyl in the atmosphere, and is rained out as dilute nitric acid. The chemistry is explored in detail in Mancinelli and Mckay (1988) and Thomas et al. (2005).

In recent years there has been a surge of evidence for moderately nearby supernovae within the last few My (Melott 2016, Wallner et al. 2016, Fimiani et al. 2016; Ludwig et al. 2016; and Binns et al. 2016). Recent assessments (Mamajek 2016; Fry et al. 2016) have favored distances close to the 40 pc proposed by Benítez, Maíz-Apellániz, and Canelles (1999). Recent studies of the effects upon the Earth of such an event have included enhanced solar ultraviolet-B (UVB 280-315 nm) due to ozone depletion, visible light effects on nocturnal organisms, increased radiation from cosmic ray secondaries (primarily muons and neutrons), and direct deposition of cosmogenic radioisotopes (Melott et al. 2017; Melott and Thomas 2018; Thomas 2018; Melott, Marinho, and Paulucci 2018).

Most biota (in the absence of humans) are nitrate-starved (Schlesinger and Bernhardt 2013) and thus it is interesting to consider the possible effects of nitrate deposition as a side of effect of astrophysical radiation enhancements. Previous assessments have suggested a small effect (Melott et al. 2005; Thomas and Honeyman 2008; Neuenswander and Melott 2015), but the longer duration of the effect of cosmic rays compared with the earlier burst situations provides an opportunity for a significant effect. Melott et al. (2017) and Thomas et al. (2016) showed that the effect of prompt ionizing photons from a supernova at 50 pc or more is quite small, three orders of magnitude lower than the threshold for major environmental effects. Therefore the focus is on the extended cosmic ray flux.

## 2. Nitrate as a possible initiator of $CO_2$ drawdown

Nitrate fertilization leading to $CO_2$ drawdown, reduced greenhouse effect, and subsequent climate cooling is not a new idea. First, terrestrial carbon uptake can easily alter atmospheric $CO_2$ levels (Le Quere et al. 2015; Keenan et al. 2016). Such carbon uptake is stimulated by deposition of additional nitrate from the atmosphere on land (Thomas et al. 2009) and in the oceans (Ganeshram et al. 1995). Nitrogen isotope changes have been documented at the Plio-Pleistocene Transition (Oesch et al. 2016) and a fertilizer-based $CO_2$ drawdown has been blamed for the late Ordovician glaciation (Shen et al 2018). The long duration of cosmic ray exposure from one or more supernovae at 50 to 100 pc motivates a re-examination of this effect, particularly since the Plio-Pleistocene cooling was accompanied by a $CO_2$ drawdown (Bartoli et al. 2011). This has been presented as a cause for the transition from woodland to savannah in Africa (Cawthra 2019), and this in turn for megaherbivore extinctions in the same area (Faith et al. 2018).

## 3. Production and deposition of nitrate from supernova enhanced cosmic ray flux

Atmospheric ionization caused by cosmic rays leads to the formation of nitrogen oxides, which catalytically deplete ozone, and are removed from the atmosphere as $HNO_3$. This leads to the deposition of $NO_3^-$ (Thomas et al., 2005). Any event that creates significant ionization of the atmosphere will cause both ozone depletion and deposition of $NO_3^-$. These production and removal processes will persist as long as the enhanced cosmic ray flux is incident on the atmosphere.

Modeling of the atmospheric chemistry effects following a nearby supernova (Melott et al. 2017) include deposition of $HNO_3$ via rain/snow. From these results, we computed a global deposition of about 7.5 Tg N per year. This is less than 10% of the preindustrial deposition of order 85 Tg per year (Schlesinger and Bernhardt 2013, p. 453). We conclude that the effect is likely to be small, although it might be greater if an increase in lightning were caused by the increased atmospheric ionization—a possibility discussed later in this paper.

## 4. Iron deposition—a possible additional fertilizer effect?

Iron is also a fertilizer, and increased iron deposition has been hypothesized as a forcing agent for terrestrial climate, and also linked to the late-Ordovician cooling (Reiners and Turchyn 2018). Iron can also act as a reducing agent for nitrates, converting them to ammonia, a more available source of nitrogen (Summers and Chang 1993). In this context we evaluate the deposition of iron, one of the major outputs of element synthesis in supernovae. Consider a supernova exploding at distance $r$ and isotropically ejecting a mass $M_{Fe}$ of iron. The resulting surface density of iron on Earth will be (Ellis, Fields, and Schramm 1996)

$$\Sigma = \frac{1}{4} \frac{f_{dust} M_{Fe}/m}{4\pi r^2} = 0.03 \; \mu\text{mol m}^{-2} \; f_{dust} \left(\frac{M_{Fe}}{0.1 \; M_\odot}\right)\left(\frac{50 \; pc}{r}\right)^2$$

Which corresponds to $9 \times 10^8$ g globally. Here $m$ is the mass of an iron atom, and the factor of ¼ is the ratio of the Earth's cross section to surface area. The factor $f_{dust} \leq 1$ accounts for the fact that at the distances of interest, only the iron that is sequestered into dust particles will traverse deeply enough into the inner solar system to arrive at Earth (Athanassiaou and Fields 1999; Fry et al 2016; Benitez et al 1999). This signal is stretched over timescales of Δt=1 $Myr$ or more (Wallner et al 2016; Ludwig et al 2016), so the average flux from the supernova is 3 x $10^{-14}$ mol m$^{-2}$ yr$^{-1}$, or about 1.7 x $10^{-12}$ g m$^{-2}$ yr$^{-1}$. Clearly, this is a tiny amount. The terrestrial flux today is about $10^{-3}$ g m$^{-2}$ yr$^{-1}$, of which no more than half is due to modern anthropogenic sources (Duce and Tindale 1991; Hamilton et al 2020). Indeed, the smallness of the supernova signal motivated a focus on live radioisotopes such as $^{60}$Fe in order to remove terrestrial background (Ellis et al 1996).

There are substantial uncertainties in the supernova distance, ejecta isotropy, iron yield, and dust fraction, at least factors of two and potentially an order of magnitude each. Even allowing for enormous favorable variations in these parameters, it is clear that the infall flux on Earth is minutely small. For comparison, the total influx of iron from extraterrestrial (and mostly interplanetary) dust onto the Earth today amounts to 0.35 $\mu$mol m$^{-2}$ yr$^{-1}$. We see that a supernova contributes a negligible fraction even of the extraterrestrial iron flux on Earth, let alone the other bioavailable iron sources. Thus, the supernova ejecta deposition will have no significant fertilizer effect.

## 5. Ozone depletion and UVB increase—a competing effect

Cosmic ray flux from a nearby supernova leads to depletion of stratospheric ozone, which results in increased penetration of solar UVB radiation to Earth's surface and into the upper part of the ocean. UVB is known to have harmful effects on most organisms, and primary producers are especially vulnerable (Llabres et al. 2013; Peng et al. 2017; Benca et al. 2018).

Atmospheric chemistry modeling of the supernova case considered here (Melott et al. 2017; Thomas 2018) show a globally averaged reduction of ozone of about 25%. Calculations of the increase in surface-level UVB and subsequent biological impacts (Thomas 2018) show a reduction in photosynthesis by marine phytoplankton ranging from a few percent to around 40%. The larger values are seen for only one biological weighting function quantifying the effects on an Antarctic phytoplankton community; two other phytoplankton inhibition weighting functions show a reduction of a few percent to around 10 percent. For a case with ozone depletion significantly higher than that considered here, Neale and Thomas (2016) found only a few percent reduction in photosynthesis for two of the most abundant, globally distributed, marine phytoplankton species.

Therefore, while reduction in primary productivity is a likely result of a nearby supernova, and would reduce the $CO_2$ drawdown effect, the impact is likely to be small.

## 6. Muon irradiation—an additional competing effect

Cosmic ray primaries generate a shower of muons, which normally constitute a substantial fraction of irradiation at the surface. In Melott et al (2017) we showed that the muon dose at the surface will reach a level of about fifteen times the total irradiation from all sources within 100 years, and remain above this normal background for thousands of years. If we add in the cosmic ray trapping effect of the magnetic field in the walls of the Local Bubble, and/or multiple supernovae (Breitschwerdt et al. 2016) the duration of the effect may be greater. Muon irradiation has the additional effect of affecting life up to a kilometer below the surface of the ocean (Melott and Thomas 2018; Melott et al. 2019). Nevertheless, only a small increase in mutation and carcinogenesis is expected for phytoplankton, although the effect may be large for megafauna (Melott et al 2019). We conclude that muon irradiation is a small competing effect with regard to atmospheric chemistry.

## 7. Lightning and Nitrate—a Possible Dominant Contributor

We have previously argued (Melott and Thomas 2019) that the large increase in atmospheric ionization (estimated as a factor of 50 in the lower troposphere) is likely to cause a major increase in the frequency of lightning. In Melott and Thomas (2019) we suggested a possible linkage to fires and tree cover changes, especially relevant for hominins in northeast Africa. In pre-industrial times, lightning is the main contributor to the fixing of atmospheric nitrogen, mostly in the form of oxides of nitrogen, which are mostly rained out as nitrate ion (Schlesinger and Bernhardt 2013). Therefore, if there is a large increase in lightning, there should be a large increase in nitrate flux as well. Our results, which so far suggest a small effect, of the order of 10%, could change completely. If the lightning enhancement is as large as suggested in Melott and Thomas (2019), this would be overwhelming. We cannot however estimate the magnitude of this effect, since there are large uncertainties in the mechanism of lighting and its connection to high energy cosmic rays.

A major increase in lightning would cause a drawdown of $CO_2$ and thereby cool the climate, helping to provide an explanation for the ongoing cooling over the Pliocene and Pleistocene. The additional flux of nitrate combined with lightning-initiated fires would enhance the transition to grasslands, as argued in Melott and Thomas (2019). More research is needed to make clear the connection between lightning and cosmic ray initiated atmospheric ionization.

## 8. Conclusions

We conclude that the fertilizer effect of nitrate deposition is likely to be small, of order 10%, for the Plio-Peistocene supernovae at a distance of 50 to 100 pc. The additional effect of fertilization by iron-bearing dust is orders of magnitude smaller. The inhibiting effects of damaging muon irradiation and UVB from ozone depletion are likely to go in the opposite direction, inhibiting photosynthesis. Therefore, for moderately nearby supernovae such as those indicated near the Plio-Pleistocene transition, it appears that no large effect is likely.

Lightning is the dominant pre-industrial source of nitrate deposition. These conclusions change completely if the high level of atmospheric ionization induces a large increase in lightning, which would cause a large increase in nitrate flux, overwhelming the other competing effects. This would enhance $CO_2$ drawdown, cooling the climate. More research is needed to clarify the cosmic ray-lightning connection. The greatest uncertainty in this assessment is the only partially understood mechanism of the initiation and propagation of lightning, and the extent of the increase in lightning, which would be caused by increased atmospheric ionization.

## 9. Acknowledgments

We thank the referee, Dr. Rocco Mancinelli, for comments which greatly improved the presentation of this manuscript.